\documentclass[preprint]{revtex4}
\usepackage{epsfig}
\begin{document}
\title{Possible CPT Violation from Planck Scale Effects}
\author
{Bipin Singh Koranga$^a$\footnote{bipinkar@phy.iitb.ac.in}, 
Mohan Narayan$^b$\footnote{mohan@udct.org} and 
S. Uma Sankar$^a$\footnote{uma@phy.iitb.ac.in}}
\affiliation
{$^a$Department of Physics 
Indian Institute of Technology, Bombay, Mumbai 400 076 \\
$^b$Department of Physics, 
Mumbai University Institute of Chemical Technology, Mumbai 400 019}

\begin{abstract}
At present there is good agreement between the neutrino 
mass-squared difference determined from the solar neutrino
data and the anti-neutrino mass-squared difference determined
from the KamLAND reactor anti-neutrino experiment. However,
the central values of the two cases differ from each other 
by about $10^{-5}$ eV$^2$. An improvement in the accuracy 
of both the solar neutrino experiments and reactor anti-neutrino
experiments can establish the existence of a non-zero difference
between neutrino and anti-neutrino mass-squared differences
and provide a signal for CPT violation. In this paper, we 
show how such a difference can arise through the CPT 
violating neutrino mass terms from Planck scale physics.
\end{abstract}

\maketitle

\section{Introduction}

KamLAND experiment recently detected the distortion due to 
oscillations in the anti-neutrino spectrum from the reactors 
and determined the corresponding mass-square difference, 
$\overline{\Delta}_{21}$,
to great precision \cite{klspec}. 
At present there is good agreement between $\overline{\Delta}_{21}$
and the mass-squared difference of the neutrinos, $\Delta_{21}$,
determined from the analysis of solar neutrino data
\cite{bahcall04}. However, the best-fit values of the two 
$\Delta$s differ from each other by $10^{-5}$ eV$^2$. 
Together, solar and KamLAND data impose the constraint $|\Delta_{21} 
-\overline{\Delta}_{21}| \leq 1.1 \times 10^{-4}$ eV$^2$
\cite{degouvea}. Future reactor 
experiments, located at a distance of about 70 Km from the 
source so that the oscillation minimum coincides with spectral 
maximum, are expected to improve the precision of 
$\overline{\Delta}_{21}$ even further \cite{srubabati}. 
Similarly future solar neutrino experiments, such as LENS
\cite{lens} and other experiments \cite{vogelaar}, 
are expected to improve the accuracy of $\Delta_{21}$.
These future experiments may indeed show that there is a
non-zero difference between $\Delta_{21}$ and $\overline{\Delta}_{21}$, 
thus establishing a signal for CPT violation in the neutrino
sector \cite{greenberg,kostelecky}. 

If $\Delta_{21}$ and $\overline{\Delta}_{21}$ are indeed found 
to be different, a natural question to ask is: How does this 
CPT violation arise? In this letter, we assume that CPT 
violation in neutrino sector arises due to Planck scale effects.
We parametrize these effects in terms of Planck scale CPT violating
neutrino mass terms and calculate the difference between $\Delta_{21}$ 
and $\overline{\Delta}_{21}$ arising due to these terms.

We assume that neutrino masses mainly arise due to Grand 
Unified Theory (GUT) dynamics via see-saw
mechanism \cite{seesaw} and these masses are CPT conserving. 
We further assume that the CPT violation arises only at 
Planck scale and parametrize it by the effective 
neutrino mass term
\begin{equation}
{\cal M_{\rm CPT}} = \frac{v^2}{2M_{\rm Pl}} \lambda_{\alpha \beta}
\label{cptvm}
\end{equation}
where $\alpha$ and $\beta$ are flavour indices.
The mass term for anti-neutrinos will have the
opposite sign \cite{barenboim}. 
Since these are effective masses arising from Planck 
scale effects, they are suppressed by $1/M_{\rm Pl}$. 
Since these are assumed to be the masses at the low energy scale, 
the electroweak vacuum expectation value (VEV) $v = 174$ GeV is 
used to make these terms have dimension of mass. In eq.~(\ref{cptvm})
the term $\lambda$ is a $3 \times 3$ matrix in flavour space whose
elements are of order 1. We assume that the Planck scale interaction 
is ``flavour blind'', 
{\em i.e.}~ the elements of the matrix $\lambda$ are independent of
$\alpha, \beta$ indices. 
In this case, the contribution to the neutrino 
mass matrix is of the form:
\begin{equation}
\mu \textstyle
\left( \begin{array}{lcr}
  1 & \!1\! & 1 \\[-.5ex]
  1 & \!1\! & 1 \\[-.5ex]
  1 & \!1\! & 1 
\end{array} \right), 
\label{textm}
\end{equation}
where the scale $\mu$ is 
\begin{equation}
\mu=v^2/2 M_{\rm Pl} \simeq 10^{-6}~\mbox{eV}. 
\label{mu}
\end{equation}
In our calculations, we take eq.~(\ref{textm}) as a perturbation to the 
main part of the neutrino mass matrix, that is generated by GUT dynamics.

\section{Calculation}

The theoretical framework, in which the Planck scale mass terms 
are treated as perturbation to GUT scale neutrino masses, is 
developed in ref. \cite{vmb}. Here we briefly recapitulate some
salient features of this framework. 
In the discussion below the labels $\alpha$ and $\beta$ refer to 
flavour eigenstates and the labels $i$ and $j$ refer to mass eigenstates.
The unperturbed ($0^{th}$-order) neutrino mass matrix ${\cal M}$ 
is diagonalized by a unitary matrix $U$ to yield the matrix
$M$, whose eigenvalues are real and non-negative.  
As stated before, ${\cal M}$ is generated
by grand unified dynamics and is related to $U$ and $M$ 
through the relation
\begin{equation}
{\cal M}=U^* \ M \ U^\dagger,\mbox{\rm where  }
 U=\left(
\begin{array}{ccc}
U_{e1} & U_{e2} & U_{e3} \\
U_{\mu 1} & U_{\mu 2} & U_{\mu 3} \\
U_{\tau 1} & U_{\tau 2} & U_{\tau 3}, 
\end{array}
\right).
\label{zordM}
\end{equation}
We adopt the usual parameterization \cite{pdg2004}:
$U_{e2}/U_{e1}=\tan\theta_{12}$, $U_{\mu3}/U_{\tau3}=\tan\theta_{23}$
and $|U_{e3}|=\sin\theta_{13}$. 
We include all possible phases in the 
definition of the neutrino mixing matrix,
\begin{equation}
U= 
{\rm diag}(e^{if_i}) \ 
R(\theta_{23}) \ 
\Pi\ 
R(\theta_{13}) \ 
\Pi^*\ 
R(\theta_{12})\ 
{\rm diag}(e^{ia_{i}}) .
\label{umatr}
\end{equation}
The phase $\delta$ appearing in 
$\Pi={\rm diag}(e^{i\delta/2},1,e^{-i\delta/2})$
is the one that enters oscillation probabilities and leads
to CP violation in neutrino oscillations.
$a_{i}$ are the so called Majorana phases and the matrix 
with this phases has the form ${\rm diag} (e^{i a_1},
e^{i a_2}, 1)$. The diagonal phase matrix on the left is given by 
${\rm diag} (e^{i f_1}, e^{i f_2}, e^{i f_3})$, which are usually 
absorbed into the definition of the respective charged lepton field. 
It is possible to rotate away the phases $f_i$, if the mass matrix
(\ref{zordM}) {\em is the complete mass matrix}.
However, since we are going to add another contribution to this
mass matrix, the phases $f_i$ of the zeroth order mass matrix 
have an impact on the complete mass matrix and thus must be retained. 
By the same token, the Majorana phases which are usually redundant
for oscillations have a dynamical role to play now.

Planck scale effects will add other contributions to the 
mass matrix. The additional term has the form given in
eq.~(\ref{textm}) and with its inclusion the neutrino  
mass matrix becomes
\begin{equation}
{\cal M}\to {\cal M}' = {\cal M}+\mu\ \lambda,
\label{eq6}
\end{equation}
It is possible that the CPT violating mass terms may be arise
not at Planck scale but at some scale $M_X$ below $M_{Pl}$ but
well above the GUT scale (by an order of magnitude or so). In
such a case, the perturbation parameter $\mu = v^2/2 M_X$, 
rather than $v^2/2 M_{pl}$. 
 
We now define the hermitian matrix 
${\cal M}'^\dagger {\cal M}'$ and 
find its eigenvalues and eigenvectors. The differences
of pairs eigenvalues give us the modified mass-squared 
differences and the eigenvectors give us the modified
neutrino mixing matrix.
To first order in the small parameter $\mu$, we have 
\begin{equation}
{\cal M}'^\dagger {\cal M}'=
{\cal M}^\dagger {\cal M}
+\mu \lambda^{\dagger} {\cal M}
+ {\cal M}^\dagger \mu\lambda.
\end{equation}
This matrix is diagonalized by a new unitary mixing matrix $U'$.
We denote this diagonal matrix, correct to first order in $\mu$,
to be $M'^2$. Using eq.~(\ref{zordM}), we can rewrite ${\cal M}$
in the above equation in terms of the diagonal matrix $M$. 
Converting ${\cal M}'$ also into its diagonal form $M'$, we 
can rewrite the above equation as 
\begin{equation}
U' {M'}^2 {U'}^\dagger =
U (M^2+m^\dagger M +M m) {U}^\dagger
\mbox{ with } m=\mu\  U^t\ \lambda\ U.
\label{tech}
\end{equation}
From the above equation, it follows that the mixing matrix $U'$,
correct to first order in $\mu$, is related to zeroth order 
mixing matrix $U$ by 
\begin{equation}
U'=U\ (1+i\delta\theta) ,
\label{eq8}
\end{equation}
where $\delta\theta$ is a hermitian 
matrix proportional to $\mu$. 

Substituting the expression for $U'$ from eq.~(\ref{eq8}) in
eq.~(\ref{tech}) we obtain
\begin{equation}
M^2+m^\dagger M +M m =
M'^2 + [ i\delta\theta, M'^2].
\label{tech1}
\end{equation}
Because $M'^2$ is diagonal, the diagonal terms of 
$[ i\delta\theta, M'^2]$ in the above equation are zero.
Therefore, to first order in $\mu$,  
the mass squared differences $\Delta M^2_{ij}=M_i^2-M_j^2$ 
get modified as:
\begin{equation}
\Delta M^{'2}_{ij} = \Delta M^2_{ij} + 2\; (M_i \mbox{Re}[m_{ii}]
-M_j  \mbox{Re}[m_{jj}]),
\label{dms}
\end{equation}
where there is no summation over the repeated indices.
The above equation shows the correction for neutrino mass-squared
difference. Because the Planck scale corrections
are assumed to be CPT violating, the correction for 
anti-neutrino mass-squared difference
will have the opposite sign.

\section{Results}
Note from eq.(\ref{dms}) that the correction term depends crucially
on the type of neutrino mass spectrum. For a hierarchial or inverse
hierachial spectrum the correction is negligible. Hence we consider
a degenerate neutrino spectrum and take the common neutrino mass
to be 2 eV, which is the upper limit from the tritium
beta decay experiment \cite{mainz}.

From the definition of the matrix $m$ in eq.~(\ref{tech}), we find
\begin{eqnarray}
m_{11} & = & \mu e^{i 2 a_1} \left( U_{e1} e^{i f_1} + U_{\mu 1} e^{i f_2}
+ U_{\tau 1} e^{i f_3} \right)^2 \nonumber \\
m_{22} & = & \mu e^{i 2 a_2} \left( U_{e2} e^{i f_1} + U_{\mu 2} e^{i f_2}
+ U_{\tau 2} e^{i f_3} \right)^2.
\end{eqnarray} 
The contribution of the terms in the Planck scale correction,
${\rm Re}(m_{22}) - {\rm Re}(m_{11})$, can be additive or subtractive 
depending on the values of the phases $a_1$ and $a_2$. Similarly the 
magnitudes of ${\rm Re}(m_{22})$ and ${\rm Re}(m_{11})$ are functions
of the phases $f_i$. Thus we try to find a combination of these
phases which can give rise to a significant difference between neutrino
and anti-neutrino mass-squared difference. In our calculations, we 
used $\theta_{12} = 34^\circ$, $\theta_{13} = 10^\circ$, $\theta_{23}
= 45^\circ$ and $\delta_{cp} = 0$. 
We define the percentage correction $P$ to be ratio of 
the difference between the corrected neutrino and anti-neutrino 
mass-squared difference to the unperturbed mass squared
difference among the first and the second mass eigenstates as;
\begin{equation}
P = 100 \times 
(\Delta M'^2_{21} - \overline{\Delta} M'^2_{21})/\Delta M^2_{21} 
\end{equation}

In Fig.(1) we plot our results as contours of constant $P$ in 
the $a_1 - a_2$ plane, for $f_i = 0$. We note that the 
maximum difference allowed is about $\pm 5 \%$ only. This is
because the magnitudes of $m_{22}$ and $m_{11}$ remain relatively
small in the limit $f_i = 0$. In Figs.(2) we plot contours
of constant $P$ in the $f_1 - f_2$ plane with $f_3 = 0 = a_1 = a_2$.
When one phase is large and the other is small, there is a possibility
that all the terms in $m_{22}$ and $m_{11}$ add up to give a large 
value for their difference and hence a large value for $P$. 
We note that a much large change, varying from $30 \%$ to $-20 \%$ 
is possible. In Figs.(3) and (4) we plot contours of constant
$P$ in $f_1 -f_3$ plane and in $f_2 - f_3$ plane. In generating
these plots, we set all other phases to zero, as in the case of Fig.(2).

The effect scales directly as the common neutrino mass, so if one takes
the WMAP \cite{tegmark} bound on the common neutrino mass which is 
0.23 eV then there is negligible effect. However if there are new flavor 
blind interactions between GUT and Planck scales, then a low common mass is
compensated by a lower scale of these new interactions and we still can 
get appreciable
effects. In fact one can even say that if the gap between $\Delta_{21}$
and $\overline{\Delta}_{21}$ persists and
future data indicate small neutrino mass, this can be an indication of new
CPT violating flavor blind interactions below Planck scale.

\begin{figure}[htb]
\includegraphics[width=\textwidth]{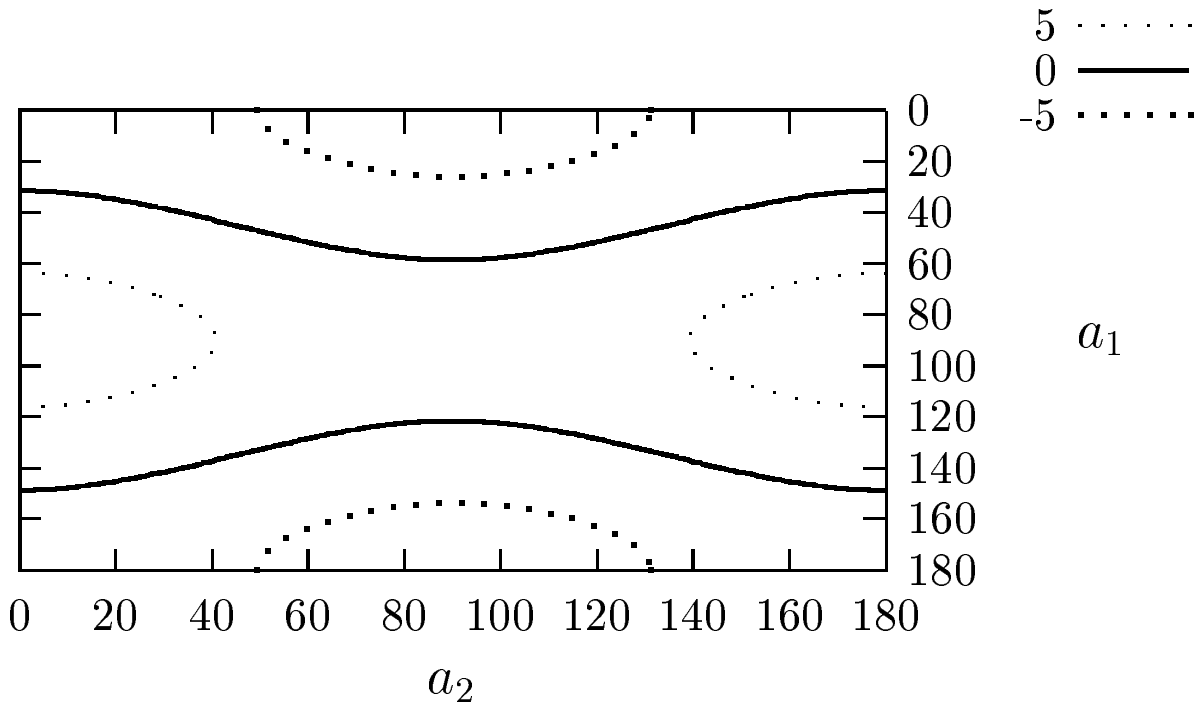}
\caption{
$P$ as a function of the Majorana phases $a_1$ and $a_2$} 
\end{figure}

\begin{figure}[htb]
\includegraphics[width=0.7\textwidth,angle=270]{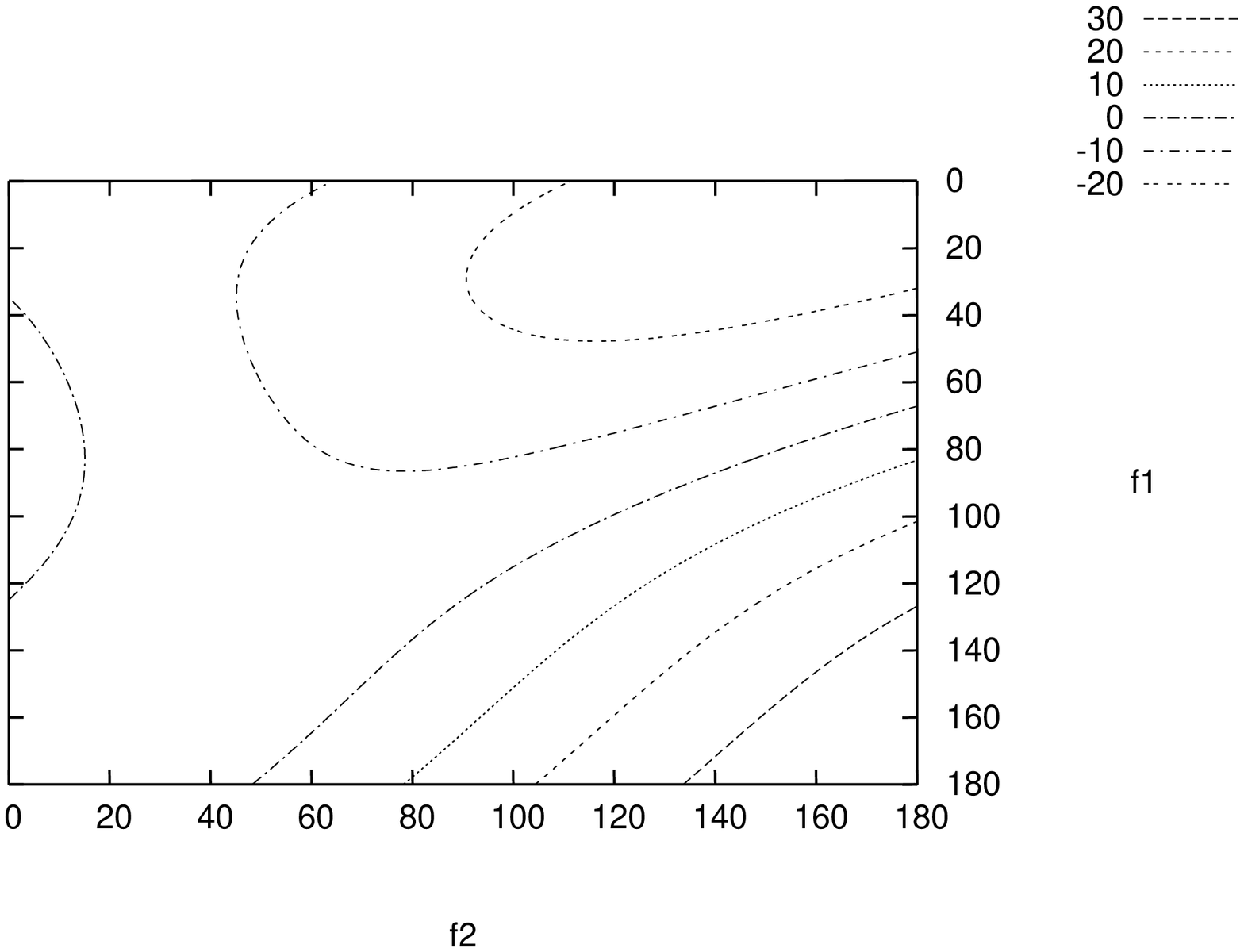}
\caption{
$P$ as a function of the phases $f_1$ and $f_2$} 
\end{figure}

\begin{figure}[htb]
\includegraphics[width=0.7\textwidth,angle=270]{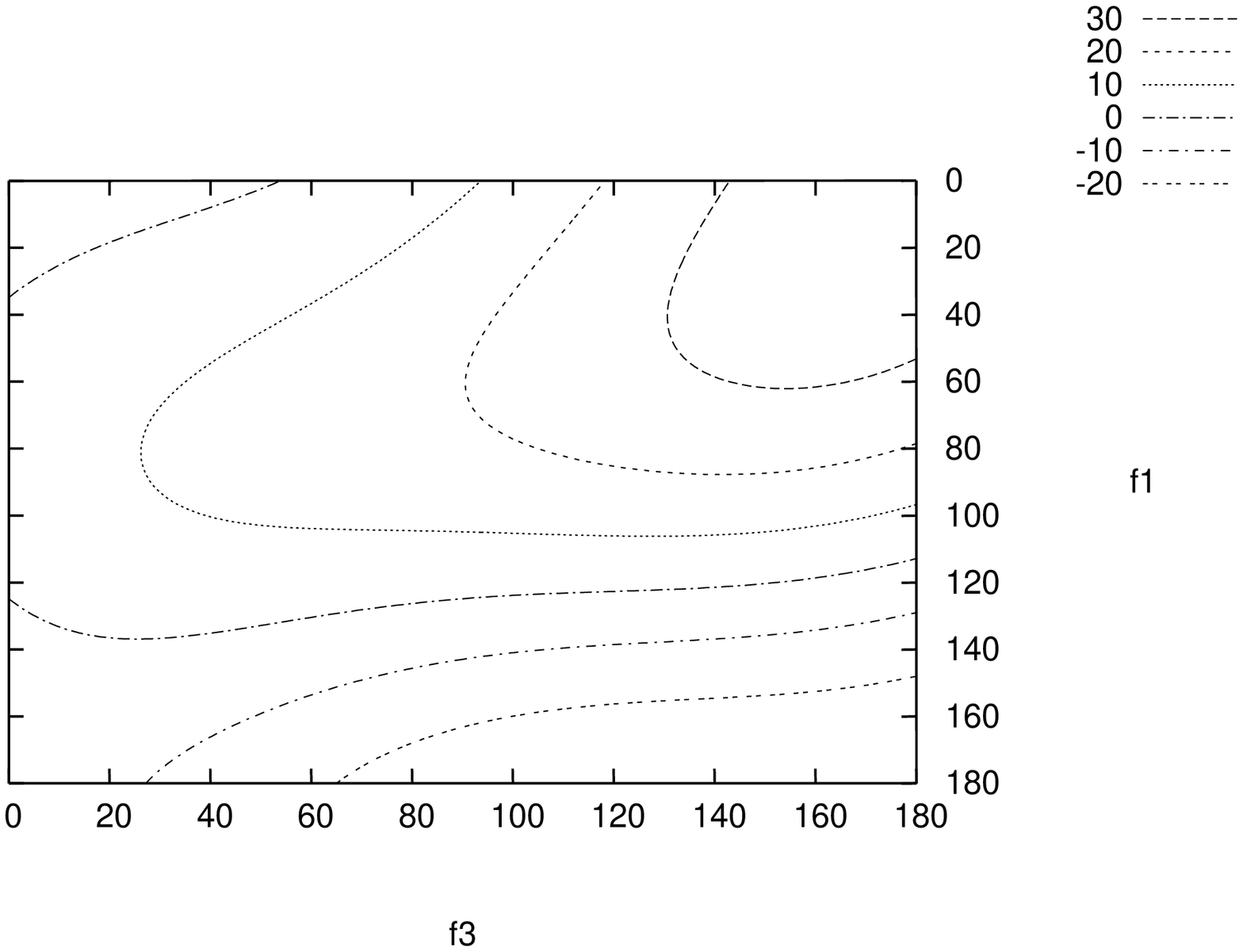}
\caption{
$P$ as a function of the phases $f_1$ and $f_3$} 
\end{figure}

\begin{figure}[htb]
\begin{center}
\includegraphics[width=0.7\textwidth,angle=270]{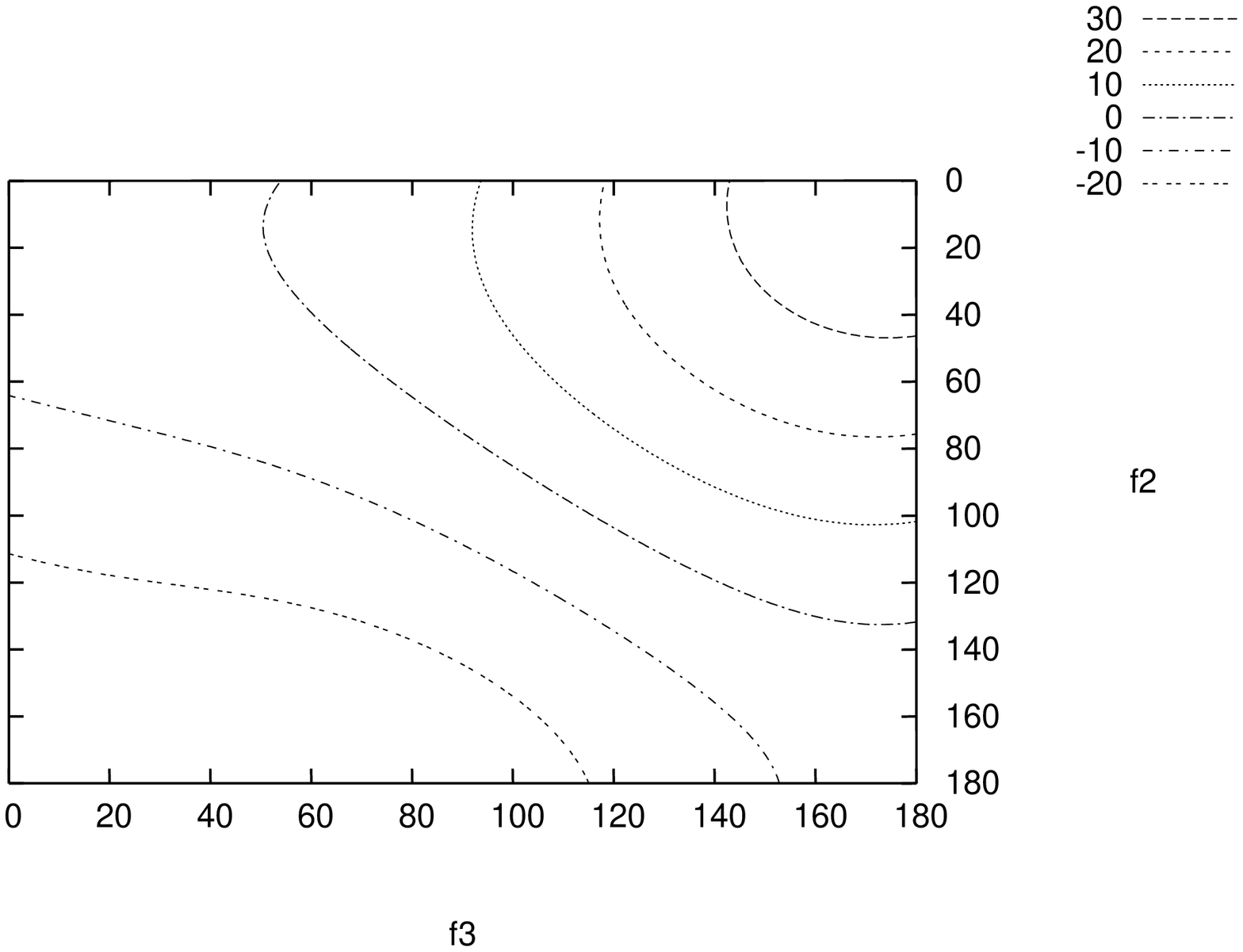}
\caption{
$P$ as a function of the phases $f_2$ and $f_3$} 
\end{center}
\end{figure}
\section{Conclusions}
Both solar and reactor data are well accounted for by invoking neutrino
oscillations. Fit to solar data give a large region for the neutrino 
mass squared difference in the two flavor parameter space. The fit to 
reactor data however gives a very strongly constrained anti-neutrino
mass squared difference. The best fits of the two mass squared differences
are appreciably different from each other. Further improvement in KamLAND
systematics and future solar neutrino data may remove this discrepancy.
However if this mismatch between the best fits persists, then CPT breaking
in the neutrino sector will be established. We have demonstrated that a 
flavour blind CPT violating neutrino masses from Planck scale physics can 
nicely accomodate this effect. 
This effect is crucially dependent on the neutrino
mass spectrum and gives rise to observable difference between 
$\Delta_{21}$ and $\overline{\Delta}_{21}$ only for a degenerate neutrino 
mass spectrum with $m_\nu \simeq 2$ eV,
which is the largest allowed value from tritium beta decay data.
The low value of the common mass implied by the WMAP bound
leads to  negligible difference between $\Delta_{21}$ and 
$\overline{\Delta}_{21}$. This can however be
compensated for by considering a slightly lower scale for the
flavour blind CPT violating mass terms rather than the usual Planck scale. 

\underline{\it Acknowledgement:} We would like to thank Francesco
Vissani for a critical reading of the manuscript.

\end{document}